\documentclass[aps,prl,reprint,amsmath,amssymb,amsfonts,floatfix, showpacs,intlimits]{revtex4-1}
\usepackage{graphicx}
\usepackage{textcomp}
\usepackage{color}
\usepackage[caption=false]{subfig}
\usepackage{color}
\usepackage[novolumeabbr]{unitsdef}
\usepackage{esint}
\usepackage{accents}
\usepackage{bbm}
\usepackage{yfonts}
\usepackage[novolumeabbr]{unitsdef}

\newunit{\dB}{dB}
\newunit{\rad}{rad}
\newcommand{\Ms}{M_s}
\newcommand{\Beff}{\vec{B}^{\mathrm{e}}}
\newcommand{\bext}{\vec{b}^{\mathrm{ext}}}
\newcommand{\bextreal}{\tilde{\vec{b}}^{\mathrm{ext}}}
\newcommand{\M}{\vec{M}}
\newcommand{\cc}{+\text{c.c.}}
\newcommand{\vmu}{\vec{\mu}}
\newcommand{\m}{\vec{m}}
\newcommand{\muz}{\mu_0}
\newcommand{\J}{\op{J}}
\newcommand{\I}{\op{I}}
\renewcommand{\P}{\op{P}}
\newcommand{\Nform}{\op{\mathcal{N}}}
\renewcommand{\vr}{\vec{r}}
\newcommand{\K}{\op{K}}
\newcommand{\dphi}{\Delta\phi}

\newcommand{\ddB}{\Delta B}

\newcommand{\cross}{\times}
\newcommand{\vgr}{\upsilon_\text{gr}}
\DeclareMathOperator{\Arg}

\renewcommand{\vec}[1]{{\boldsymbol{#1}}}
\newcommand{\op}[1]{\hat{\boldsymbol{#1}}}

\begin{document}

  \title{Bias-free spin-wave phase shifter for magnonic logic}

  \author{Steven Louis}
  \affiliation{Department of Physics, Oakland University, 2200 N. Squirrel Rd., Rochester, Michigan,
      48309--4401, USA}

  \author{Ivan Lisenkov}
  \email[]{ivan.lisenkov@phystech.edu}
  \affiliation{Department of Physics, Oakland University, 2200 N. Squirrel Rd., Rochester, Michigan,
      48309--4401, USA}
  \affiliation{Kotelnikov Institute of Radio-engineering and Electronics of RAS, 11--7 Mokhovaya st.,
      Moscow, 125009, Russia}

  \author{Sergey Nikitov}
  \affiliation{Kotelnikov Institute of Radio-engineering and Electronics of RAS, 11--7 Mokhovaya st.,
  	Moscow, 125009, Russia}
  \affiliation{Moscow Institute of Physics and Technology, 9 Instituskij per., Dolgoprudny, 141700, Moscow
  	Region,  Russia}

  \author{Vasyl Tyberkevych}
  \affiliation{Department of Physics, Oakland University, 2200 N. Squirrel Rd., Rochester, Michigan,
      48309--4401, USA}

  \author{Andrei Slavin}
  \affiliation{Department of Physics, Oakland University, 2200 N. Squirrel Rd., Rochester, Michigan,
      48309--4401, USA}

  \date{\today}

  \begin{abstract}
     A design of a magnonic phase shifter operating without an external bias magnetic field is proposed. The phase shifter uses a localized collective spin wave mode propagating along a domain wall ``waveguide'' in a dipolarly-coupled magnetic dot array existing in a chessboard antiferromagnetic (CAFM) ground state. It is demonstrated numerically that remagnetization of a single magnetic dot adjacent to the domain wall waveguide introduces a controllable phase shift in the propagating spin wave mode without significant change of the mode amplitude. It is also demonstrated that a logic XOR gate can be realized in the same system.
  \end{abstract}

\maketitle

The emerging field of magnonics~\cite{bib:Chumak:2015, nikitov2015magnonics, Chumak2014Magnon, khitun2010magnonic, demokritov2012magnonics} represents a vehicle to implement novel computing architectures including neuromorphic~\cite{bib:Sengupta:2016, bib:Nikonov:2015}, holographic~\cite{bib:Kozhevnikov:2015, bib:Gertz:2016}, and interference-based~\cite{Fetisov1999Microwave, Chumak2014Magnon} computing. In these new computing methods, information can be encoded in the phase rather than in the amplitude of the carrier signal. With phase-encoded signals, logic inversion is implemented by the addition of $\pi$ to the signal phase (\emph{phase inversion}). Ideally, phase inversion in a magnonic system should work in a wide frequency bandwidth, with minimal attenuation, and without an external bias magnetic field.

Progress has been made in bringing magnonic devices from theory to application. Studies have demonstrated magnonic transistors~\cite{Chumak2014Magnon}, logic devices~\cite{ bib:Khitun:2010, Fetisov1999Microwave, Schneider2008Realization, *Serga2009Generation, Nikitin2015A, Klinger2015Spin, bib:Ding:2012}, waveguides~\cite{Pirro2014Spin, rychly2015magnonic,garcia2015narrow}, and phase shifters~\cite{Hansen2009Dual, Ustinov2014Nonlinear, bib:Kostylev:2007}. Also, precise manipulation of spin waves has been demonstrated using magnetic fields induced by electric currents adjacent to ferromagnetic films~\cite{bib:Demidov:2009, bib:Kostylev:2007, bib:Rousseau:2015, bib:Vasiliev:2007, bib:Chumak:2010, vogt2012spin, *bib:Vogt:2014}. Unfortunately, these magnonic systems require the use of external bias magnetic fields and electric currents, thus making miniaturization and pairing of magnonic devices with CMOS-type integrated circuits rather difficult.

Thus, \emph{self-biased} magnonic systems are highly desirable. This can be achieved with nanostructured magnetic materials with discreet magnetic elements, such as arrays of magnetic nanodots~\cite{bib:Gubbiotti:2004, *bib:Tacchi:2010, *bib:Tacchi:2011, *bib:Tacchi:2011a, bib:Kruglyak:2010, bib:Zivieri:2011, bib:Carlotti:2014}. By using magnetic elements that are small enough to prevent formation of multiple magnetic domains, magnetic nanodot arrays can retain their magnetic ground state without an applied magnetic field. Also, arrays of magnetic elements have an additional degree of freedom: elements within the array have multistable magnetization directions, making it possible to \emph{dynamically engineer} quasi-stable static magnetic states of arrays by adjusting magnetization direction of individual elements and, consequently, changing the properties of collective spin wave excitations of an array~\cite{bib:Ding:2012, verba2012collective, bib:Verba:2012a,bib:Goolaup:2007, bib:Haldar:2016a}.

Recent experiments~\cite{bib:Stebliy:2015, bib:Haldar:2016a} indicate that a self-biased magnetic array of discrete, dynamically controllable elements can provide a medium where spin waves can propagate, which then can be controlled by the \emph{local} alteration of the magnetic state of individual array elements, or in other words, by introducing defects in the quasi-stable static magnetization state of an array~\cite{lisenkov2015theoretical}. This can be achieved not only by applying an external non-uniform field~\cite{vogt2012spin}, but also by using the effect of voltage controlled magnetic anisotropy~\cite{bib:Wang:2011, bib:Matsukura:2015}

In this work, performed by means of numerical simulations, we demonstrate two ideas: first, that a~\emph{domain wall} separating two regions in a magnetic dot array, existing in the chessboard antiferromagnetic (CAFM) ground state, can convey a highly localized collective spin-wave mode, thus serving as a spin-wave ``waveguide'', and, second, that the creation of a \emph{point defect} near the domain wall waveguide(e.g. by remagnetization of a single magnetic dot adjacent to the waveguide) induces a controllable phase shift in the propagating spin wave mode. We found that the induced phase shift can be as high as $\pi$, and is nearly constant in a wide frequency band ($\approx400\MHz$ in our simulations). At the same time, the additional losses of the spin wave amplitude caused by the point defect can be as low as \ilu[1.0]{\dB}. As a possible application, we have also demonstrated a magnonic exclusive disjunction (XOR) gate by organizing domain wall waveguides into a scheme of a Mach-Zehnder interferometer~\cite{bib:Rousseau:2015, Schneider2008Realization, bib:Ding:2012} with a point defect in each arm, where the states of the defects serve as logic inputs and the amplitude of the spin-wave signal after the interference serves is the output.

\begin{figure}
\centering
\includegraphics[width=\linewidth]{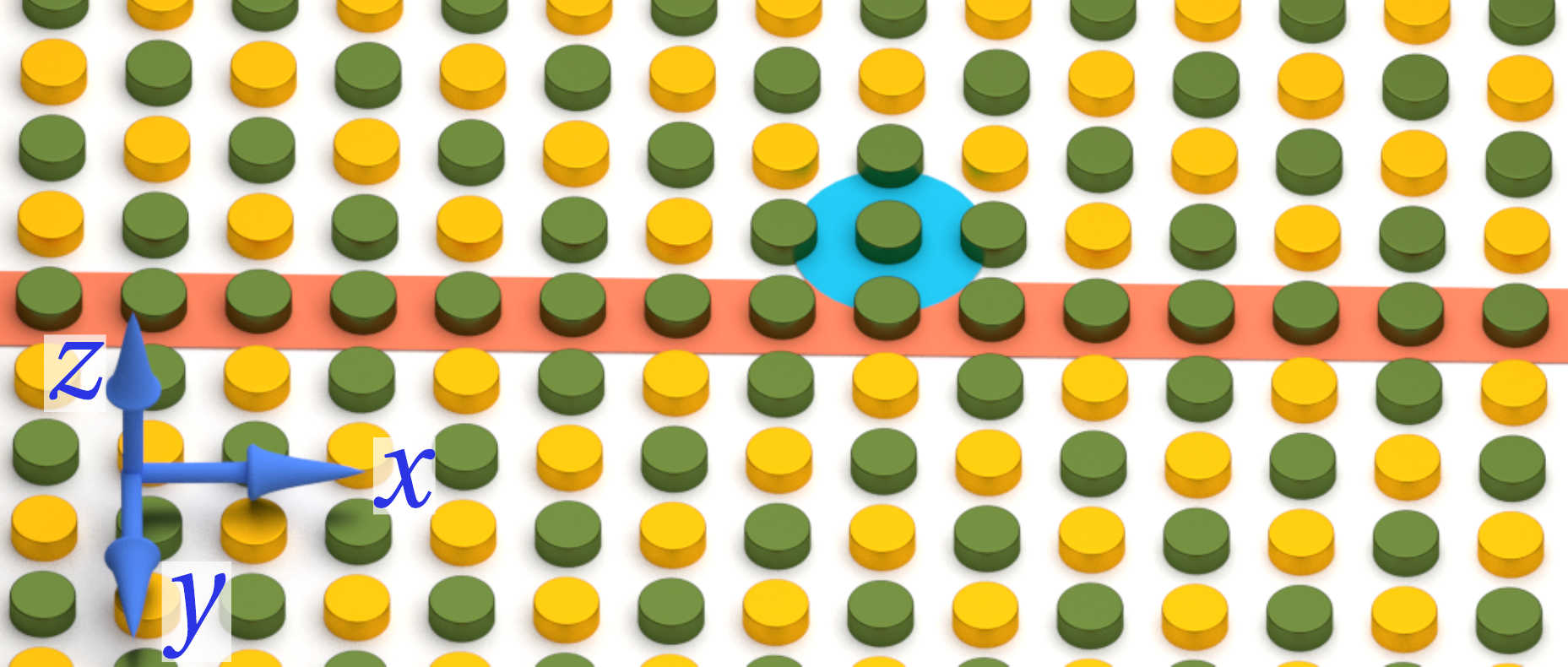}
\caption{An example of a nanodot array with a domain wall waveguide (highlighted in red) and a point defect (highlighted in blue) adjacent to the waveguide. Green (yellow) dots have static magnetization pointing into (out of) the page.}
\label{fig:invGS}
\end{figure}

We consider a nano-strucutured magnetic meta-material based on an array of magnetically saturated cylindrical pillars (nanodots) with uniform magnetization. The nanodots are arranged periodically in a square lattice and are coupled by dipolar interaction. The shape of the dot and its perpendicular uniaxial magnetic anisotropy are chosen in such a way, that in equilibrium, the magnetization of each dot is perpendicular to the array plane (pointing either upwards or downwards). The ground state of such an array is a \emph{chessboart antiferromagnetic} (CAFM)state, with the magnetizations of neighboring nanodots aligned in the opposite directions, see Fig.~\ref{fig:invGS}. The CAFM ground state is stable in the absence of externally applied bias magnetic field, and has zero net magnetic moment~\cite{bib:Verba:2012a}.

\begin{figure}
\centering
\includegraphics{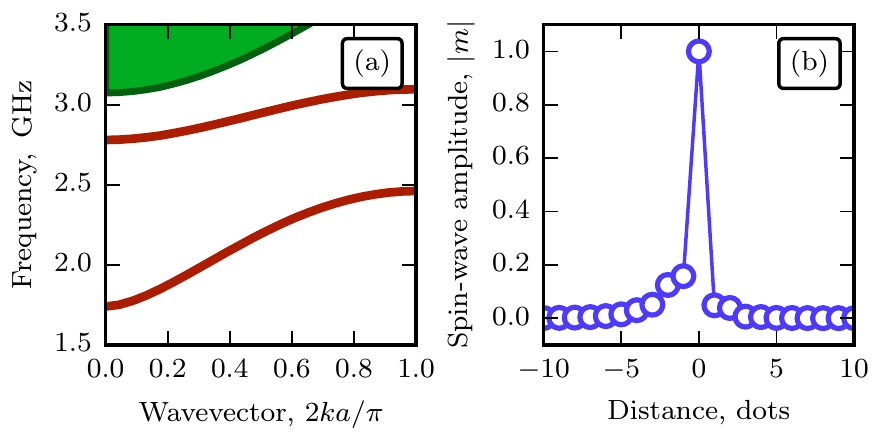}
\caption{(a) Dispersion of spin waves in an magnetic dot array existing in a CAFM ground state and containing a domain wall waveguide. The red lines show the dispersion of spin wave modes localized on the waveguide, while the green area shows the bulk spin wave spectrum of the array; (b) Distribution of the dynamic magnetization in the direction perpendicular to the waveguide in a localized spin wave mode. Array parameters: dot radius \ilu[30]{\nm}, dot height \ilu[60]{\nm}, distance between the centers of adjacent dots \ilu[90]{\nm}, saturation magnetization \ilu{800}{\kiloampere/\meter}, and the energy of the perpendicular magnetic anisotropy \ilu[0.25]{\megajoule/\cubicmeter}.}
\label{fig:dispersion}
\end{figure}

For an array existing in a CAFM ground state, the domain wall is a line defect in the form of a series of neighboring dots with static magnetization pointing in the same direction (see the line highlighted in red in Fig.~\ref{fig:invGS}). This domain wall creates a local minimum in the internal magnetic field inside the array (local potential well), which supports the formation and propagation of highly-localized spin wave modes~\cite{lisenkov2015theoretical, xing2015waveguide} with frequencies well separated from the bulk spin wave spectrum of the CAFM array (see Fig.~\ref{fig:dispersion}).

We performed numerical simulations for this system using the ``macrospin'' approximation, which assumes that each dot is uniformly magnetized, and that the dominant precession mode is uniform~\cite{verba2012collective, lisenkov2015theoretical}. Here we are mostly interested in dynamics of \emph{forced} spin waves, excited by a point-like harmonic magnetic field. This type of excitation can model, for example, spin-torque nano-oscillators~\cite{Slavin2009Nonlinear} embedded in the array.

\begin{figure*}
\centering
\includegraphics{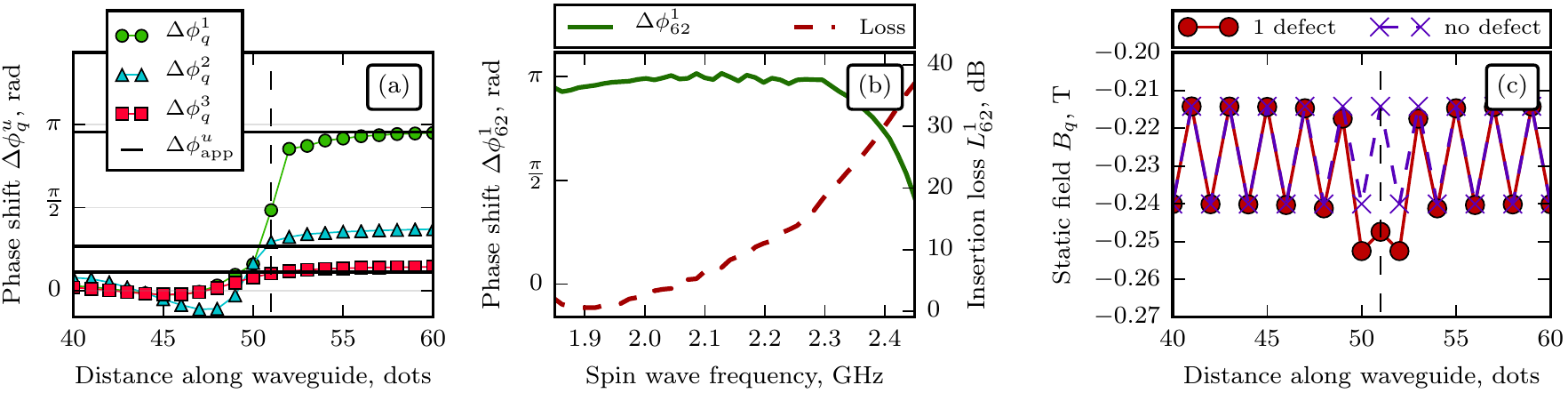}
\caption{(a) Phase shift $\dphi_p^u$ of the waveguide spin wave mode induced by a local point defect as a function of the position $p$ along the waveguide.
The index $u$ denotes the vertical position of the point defect above the waveguide: $u$=1 means that defect is directly adjacent to the waveguide, $u$=2,3 means that defect is two (three) dots away from the waveguide. The dashed vertical line shows the horizontal position of the point defect (above the dot $q$=51). The horizontal solid black lines show the values of the phase shift calculated using the approximate analytic formula~\eqref{eq:approx_phase_shift}; (b) Phase shift $\dphi_{62}^1$ and insertion losses $L_{62}^1$ introduced by an adjacent ($u=1$) point defect as a function of the excitation frequency; (c) Distribution of the effective internal magnetic field $B_q$ (\ref{eq:static}) along the waveguide in the presence (red solid line) and absence (dashed line) of the point defect. Parameters of the array are the same as in Fig.~\ref{fig:dispersion}.}
\label{fig:beef}
\end{figure*}

The dynamic magnetization for each nanodot can be described by the Landau-Lifshitz equation:
\begin{equation}
 \dfrac{d\M_j}{dt} = \gamma(\Beff_j\times\vec\M_j) + \gamma(\bextreal_j\times\M_j) + \alpha_G \M_j\times\dfrac{d\M_j}{dt},
 \label{eq:LL}
\end{equation}
where $\vec{M}_j$ is the magnetization of the $j$-th dot, $\Beff_j$ is the effective dipolar field acting on the $j$-th dot by all other dots in the array, $\bextreal_j$ is the external excitation field, $\alpha_G$ is the Gilbert damping constant and $\gamma/(2\pi) \approx 28{\GHz}/{\tesla}$ is the modulus of gyromagnetic ratio.
The diploar field acting on the $j$-th dot is calculated as $\Beff_j =-\mu_0\sum_l\op{N}_{jl}\cdot\vec{M}_l$, where $l$ indexes all dots in the array. Here $\mu_0$ is the magnetic constant and $\op{N}_{ij}$ is the effective demagnetization tensor~\cite{lisenkov2015theoretical}: $\op{N}_{jl} =  \Nform(\vr_j - \vr_l) + \delta_{jl}\K$
where $\Nform(\vr)$ is the shape demagnetization tensor of a dot~\cite{verba2012collective, bib:Tandon:2004}, $\vr_j$ is a position vector of $j$-th dot,
$\K = -K^a/(2\muz\Ms^2)\, \vec{n}\otimes\vec{n}$, $K^a$ is the energy of the uniaxial magnetic anisotropy, $\vec{n}$ is a unit vector perpendicular to the array plane, and $\Ms = |\M_j|$.

The harmonic excitation by a point-like magnetic field can be represented as:
$\bextreal_j = \bext_j e^{-i\omega t}  \cc$, where $\bext_j$ is the complex amplitude of the excitation, $\omega = 2\pi f$ is the excitation frequency. Under this excitation, the magnetization $\vec{M}_j$ of each dot will have a time dependence, which can be split into static and dynamic parts:
\begin{equation}
\vec{M}_j=M_s\left(\vec{\mu}_j+(\vec{m}_j  e^{-i \omega t} \cc) \right),
\label{eq:split}
\end{equation}
where the unit vector $\vmu_j$ is the direction of static magnetization and $\m_j$ is the small dimensionless deviation of magnetization of the $j$-th dot. Note that $\m_j\cdot\vmu_j=0$ as these vectors are orthogonal .

Substituting the expansion \eqref{eq:split} into \eqref{eq:LL} and omitting second order terms of $\m_j$, we can express \eqref{eq:LL} as two separate static and dynamic equations:
\begin{subequations}
    \begin{gather}
        B_j=-\muz \Ms \vmu_j\cdot\sum_l\op{N}_{jl}\cdot\vmu_l\label{eq:static},\\
        \begin{aligned}
        -i\omega\J_j\cdot\m_j - \left(\sum_l\op{\Omega}_{jl} + i\alpha_G\omega\I\delta_{jl}\right)\cdot\m_l = \gamma\P_j\cdot \bext_j,
        \end{aligned}
        \label{eq:dynamic}
    \end{gather}
    \label{eq:main}
\end{subequations}
where $B_j$ is the scalar static demagnetization field of the $j$th dot, $\J_j = \vec{e}\cdot\vmu_j$, $\vec{e}$ is the Levi-Civita symbol~\cite{bib:Riley:2006}, $\P_j = - \J_j^2$ is the projection operator~\cite{lisenkov2015theoretical},
$\op{\Omega}_{jl}=\gamma (B_j\delta_{jl}\op{I} + \gamma\mu_0 M_s\op{P}_j\cdot\op{N}_{jl}\cdot\op{P}_l)$ is the dynamic interaction tensor~\cite{verba2012collective}, and $\op{I}$ is an identity matrix. The static equation~\eqref{eq:static} can be easily solved, because in our system the ground state of the array (distribution $\vmu_j$) is defined as CAFM, and the dots have only two stable static magnetization directions ($\vmu_j =\pm \vec{n}$). The system~\eqref{eq:dynamic} represents a linearized version of~\eqref{eq:LL}, written as a generalized inhomogenious system of linear equations~\cite{bib:Buijnsters:2014, lisenkov2015theoretical}.

Let us, first, consider a domain wall (or a line defect) in the CAFM ground state of the array, and show that it forms a spin wave waveguide, see Fig.~\ref{fig:invGS}. To find the elementary spin wave solutions propagating along the waveguide, we introduce the dot index as $j = (p,q,n)$, where $p$ and $q$ indices denote the position of a unit cell of the array along the directions that are parallel and perpendicular to the waveguide, respectively, and the index $n$ denotes the position of the dot inside a unit cell. The dynamic dot magnetization $\m_j=\m_{(p,q,n)}$ can be written as:$
    \m_{(p,q,n)} = \m_{(p,n)} e^{2ikaq}$
where $a$ is the distance between the dots in a square lattice and $k$ is the wavevector of the spin wave mode localized on the domain wall waveguide. Substituting the dynamic magnetization in ~\eqref{eq:dynamic}, neglecting the excitation and damping terms, and solving the obtained homogeneous eigen-value problem, one can find the spectrum of spin-wave eigen-excitations (see Section \textbf{V} in~[\onlinecite{lisenkov2015theoretical}] for technical details).

The dispersion relation calculated for a CAFM array with a domain wall waveguide is presented in Fig.~\ref{fig:dispersion}(a). The parameters of the array are chosen as follows~\cite{Nekrash2016}: pillar radius, $r=30\nm$, pillar height $h=60\nm$, distance between centers of the nearest neighbors $a=90\nm$, saturation magnetization $\Ms = 800\kiloampere/\meter$, energy of the perpendicular magnetic anisotropy $K_a = 0.25 \megajoule/\cubicmeter$. The array has two types of modes: localized waveguide modes (dispersion curves plotted in red) and bulk modes (spectrum plotted in green). The waveguide modes are localized (or trapped) inside the potential well formed by the nonuniform internal static dipolar field in the vicinity of the domain wall waveguide~\cite{bib:Jorzick:2001, lisenkov2014spin, lisenkov2015theoretical}. In our case the waveguide spin-wave mode is strongly localized and its amplitude drops significantly a few dots away from the waveguide (see Fig.~\ref{fig:dispersion}(b)). The profile of the spin wave is slightly asymmetric due to a non-zero net magnetic moment of the defect line and corresponding Damon-Eshbach-type symmetry breaking~\cite{lisenkov2014spin, bib:Eshbach:1960}.

A point defect, created by reversing the magnetization of one of the dots near the waveguide, will break the translational symmetry of the system. To understand the impact of a single point defect, we used \eqref{eq:main} to simulate an array of 109 by 25 dots, with a waveguide in the central (13th) row of dots. The point defect was created above the 51st dot along the waveguide with a certain offset $u$ from the waveguide ($u=1$ corresponds to the defect just above the waveguide, as it is shown in Fig.~\ref{fig:invGS}). The spin wave mode propagating along the waveguide was excited by a point source located at the beginning of the waveguide.

To investigate how the point defect affects the propagation of the localized spin waves we calculated the dynamic parts of the dot's magnetization along the waveguide in the presence and absence of the point defect, $\m^u_q$ and $\m^0_q$, respectively ($q$ is the dot index along the waveguide and $u$ is the offset of the defect dot from the waveguide). Using these values we calculated the ``insertion loss'' $L_q^u$ and phase shift $\dphi_p^u$ caused by the defect as:
\begin{gather}
    L_p^u = 10\log(|\m^0_p|^2/|\m^u_p|^2),\\
    \dphi_p^u = \phi_p^u - \phi_p^0,
\end{gather}
where the spin-wave phases are calculated as: $\phi_{p} = \arg\left(-i \vmu_p\cdot\left[\left(\m_p\right)^*\cross\m_0\right]\right)$ and $\m_0$ is the magnetization at the point of excitation. As it follows from the above definitions, in the absence of any effect ($\m_q^u \to \m_q^0$) $L_q^u\equiv0$ and $\dphi_q^u\equiv0$.

The defect-induced phase shift $\dphi_q^u$ as a function of the dot position $q$ along the waveguide for defect vertical offsets of $u=1,2,3$ is shown in Fig.~\ref{fig:beef}(a). The excitation frequency was \ilu[1.9]{\GHz}. In the region between the source and the defect ($q<51$) the presence of the defect has practically no influence on the phase of the propagating spin wave. As the spin wave passes the defect ($q>51$), it acquires a finite phase shift $\dphi^u$ and continues to propagate along the waveguide. For a point defect that is $u=1$ dot offset from the waveguide (as in Fig.~\ref{fig:invGS}), the phase shift is $\dphi^1\approx0.95\pi\rad$.

When the point defect is located farther from the waveguide, its influence, obviously, decreases: $\dphi^2\approx\pi/3\rad$ and $\dphi^3\approx2\pi/5\rad$. For defects located even further from the waveguide we registered progressively smaller induced phase shifts.

The defect-induced phase shift for the offset $u=1$ at the dot number $q=62$ is plotted in Fig.~\ref{fig:beef}(b) as a function of spin wave frequency. The frequency range is selected within the propagation band of the first localized spin wave mode (see Fig.~\ref{fig:dispersion}). The main feature of the frequency dependence of the phase shift is its flatness: the phase shift $\dphi^1$ remains constant in a wide frequency region (approximately \ilu[400]{\MHz}). At the same time, the insertion losses $L^1$ increase substantially with the frequency, but near the lower end of the spin wave frequency band ($f\approx\ilu[1.9]{\GHz}$) they are quite low, and do not exceed \ilu[1]{\dB}.

The defect-induced phase shift in the propagating spin wave is caused, primarily, by the changes in the internal static dipolar field~\cite{bib:Vasiliev:2007, bib:Kostylev:2007, bib:Rousseau:2015}. Fig.~\ref{fig:beef}(c) shows the profile of the internal field $B_q$ along the waveguide for two cases: in the presence and absence of the point defect with an offset $u=1$. For the case without a defect, the internal field varies periodically. The defect interrupts this perfect periodicity, causing a significant change in the local profile of the internal magnetic field. Roughly, this change in the internal magnetic field locally modifies the dispersion relation of the waveguide spin wave mode, thus lowering its frequency.~\cite{bib:Kostylev:2007}.

\begin{figure*}\centering
\includegraphics{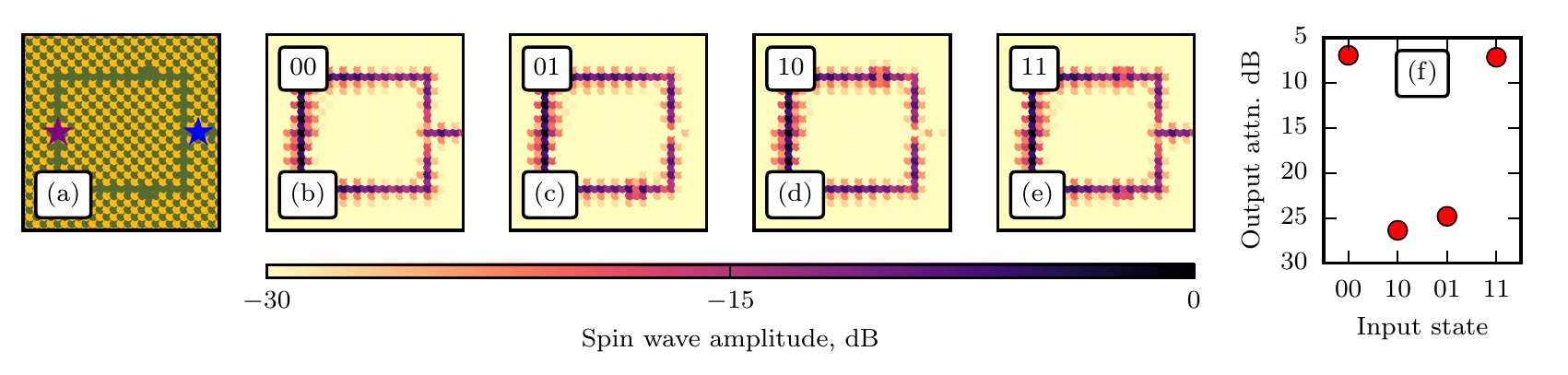}
\caption{An example of a magnonic XOR logic gate implemented using the scheme of a Mach-Zehnder interferometer. (a) The ground state of the array in the {\tt 11} logic state. Green (yellow) dots have the magnetization pointing down (up). Purple and blue stars are, respectively, the point of excitation and the output; (b-e) Distribution of the spin wave amplitude in the gate for different logic inputs. (f) Amplitude of the spin wave at the output for different logic inputs, normalized by the excitation power.  Parameters of the array are the same as in Fig.~\ref{fig:dispersion}, frequency of excitation \ilu[1.9]{\GHz}\label{fig:xor}.}
\end{figure*}

This explains why the spin waves excited near the upper boundary of the frequency interval of existence of the waveguide spin wave mode are strongly attenuated by the presence of the defect (see Fig.~\ref{fig:beef}(b)): the spin waves with such frequencies encounter a local band gap created near the defect and cannot propagate through this area. At the same time, the spin waves with lower frequencies can propagate though the ``defect'' area, but their dispersion in this region is modified and, consequently, they experience an additional phase shift.

Dispersion of the waveguide dipolar spin wave mode is practically linear: $\omega \approx \omega_0 + \vgr k$, where
$\vgr\approx1.1\kilometer/\second$ is the spin wave group velocity, see Fig.~\ref{fig:dispersion}(a). The difference in the static field for each dot changes the ``local'' spin wave eigenfrequency by $\omega_0$ by $\Delta\omega_{0q}^u=\gamma\ddB_q^u$, which, in turn, changes the local wavenumber at a given excitation frequency by $\Delta k^u_q = \Delta\omega_{0q}^u/\vgr$. Thus, the total phase shift accumulated by the wave passing though the defect region is:
\begin{equation}
    \dphi^u_{\text{app}} \approx \dfrac{\gamma a}{\vgr} \sum_q \ddB_q^u,
    \label{eq:approx_phase_shift}
\end{equation}
where the sum is taken over all the dots in the waveguide.
The dots located near the defect experience an internal field that is reduced by $\ddB^u_q= 2 \muz\Ms \op{N}^{zz}(\vr_q - \vr_u)$, where $\vr_u$ is the radius vector of the defect. This sum can be evaluated by calculating the Fourier transform of the demagnetization tensor: $\sum_q \ddB_q = 2 \muz\Ms\op{E}^{zz}_0(u)$ (see (34) in~[\onlinecite{lisenkov2015theoretical}]). The acquired additional phase shift calculated using~\eqref{eq:approx_phase_shift} and the results of the direct numerical simulations are compared in Fig.~\ref{fig:beef}(a)

As an example of application of the above described dynamically reconfigurable magnonic phase shifter we consider an interferometer-based magnonic exclusive disjunction (XOR) logic gate. The domain wall waveguides in the CAFM ground state of the dot array form the arms of the Mach-Zehnder interferometer, while the two point defects in different arms of the interferometer serve as the logical inputs of the gate, see Fig.~\ref{fig:xor}(a). When a point defect is engaged, we consider this as the logic state {\tt 1}, while an absence of a defect represents the logic state {\tt 0}. The spin waves are excited at the frequency of \ilu[1.9]{\GHz} by a point-like magnetic field located at the place shown by the purple star. The logic output of the XOR gate is simply the amplitude of the resulting spin wave at the exit of the interferometer (blue star in~Fig.~\ref{fig:xor}(a)).

When both defects are not engaged (logic state {\tt00}) spin waves traveling in both arms of the interferometer accrue the same phase, the interference is constructive, and the output spin wave amplitude is high, see Fig.~\ref{fig:xor}(b). If only one of the defects is engaged (logical states {\tt01} and {\tt10}, see Fig.~\ref{fig:xor}(c,d)), the spin wave in one of the arms accrues an additional phase shift of $\pi$ and the interference is destructive at the output port, providing a low output amplitude. Finally, if both defects are engaged, spin waves in \emph{both} arms receive the same phase shift of $\pi$, which restores the constructive interference and high output level, see~Fig.~\ref{fig:xor}(e). In our case, the difference in the output signal amplitudes was more than \ilu{>15}{\dB} (see Fig.~\ref{fig:xor}(f) for comparison of the signal outputs for different logic input states), which allows one to easily introduce a power threshold for the output logic state.

In summary, arrays of magnetic nanodots are a promising platform for future magnonic circuitry. Here we have demonstrated a possibility of fine control of spin wave propagation by a local modification of the ground state of a nanodot array. The domain wall ``waveguides'' (line defects) can be used to guide spin waves, while dynamically reconfigurable point defects can locally modify the dispersion of the waveguide spin wave modes. It was demonstrated that a point defect can serve as an efficient phase shifter that introduces a fixed phase shift over a wide range of spin wave frequencies and has low insertion losses. For the parameters of our simulations the phase shift caused by a single point defect can be as high as $\pi$. The value of the phase shift can be controlled by the distance between the defect and the waveguide. As an application of the defect-based phase shifter we demonstrated a magnonic XOR gate, where the states of the two defects serve as logic inputs, and the amplitude of the resultant spin wave serves as a logic output. The difference in the output amplitude for {\tt 0} and {\tt 1} output state can be very high (\ilu[>15]{\dB}), which makes the proposed magnonic XOR gate practically viable.

\begin{acknowledgments}
This work was supported in part by the CNFD grant from the Semiconductor Research Corporation, by the Grant ECCS-1305586 from the National Science Foundation of the USA, by the contract from the US Army TARDEC, RDECOM, and by the DARPA grant ``Coherent Information Transduction between Photons, Magnons, and Electric Charge Carriers''. I.L. and S.N. acknowledge the financial support from the Russian Scientific Foundation under the Grant \#14-19-00760.
\end{acknowledgments}

\bibliography{FMwaveguide}


\end{document}